\documentclass[a4paper]{jpconf}
\usepackage{geometry,amsmath,amsfonts}
\usepackage{slashed}
\usepackage{epsfig}
\usepackage{latexsym}
\usepackage{graphicx}
\usepackage{epstopdf}  % to use eps with Typeset -> Pdftex
\usepackage[justification=RaggedRight]{caption}
\usepackage{amssymb}
\usepackage{subfig}
\usepackage{color}
\usepackage{multirow}
\usepackage{color}
\usepackage{rotating}
\usepackage{ifthen}
\usepackage{epsfig}
\usepackage{graphics}
\usepackage{xcolor}
\usepackage{enumitem}
\begin{document}
\title{Dark Matter phenomenology of GUT Inspired simplified models}

\author{Giorgio Arcadi}

\address{Laboratoire de Physique Th\'{e}orique, CNRS, Univ. Paris-Sud, Universit\'{e} 
Paris-Saclay, 91405 Orsay, France}

\ead{giorgio.arcadi@th.u-psud.fr}

\begin{abstract}
We discuss some aspects of dark matter phenomenology, in particular related to Direct detection and collider searches, of models in which a fermionic Dark Matter interacts with SM fermions through spin 1 mediators. Contrary to conventional simplified models we will consider fixed assignments of the couplings of the ($Z^{'}$) mediator, according theoretically motivated embeddings. This allows to predict signals at future experimental facilities which can be used to test and possibly discriminate different realizations. 
\end{abstract}

%Preprint:
{\bf Preprint number:} LPT-Orsay-15-83

\section{Introduction}

Simplified Dark Matter models, namely simple extensions of the SM featuring a WIMP dark matter candidate and a mediator of its interactions with SM states, have proven to be a very powerful tool for facing the outcome of current dark matter and collider related searches. They are typically built following a bottom-up approach and are characterized by a limited set of free parameters including the DM and mediator masses as well as the coupling of the mediator with the DM and the SM states. The aim of this work is to investigate some possible advantages of embedding these kinds of setups in more theoretically motivated frameworks. These are twofold. First of all it is possible, in this way, to further reduce the amount of free parameters, assigning definite values to the couplings of the theory. In addition, being predictions of the couplings of theory available, it is possible to infer hypothetical next future signals which can be probed at next future DM detection facilities or at the LHC run II. We also remark that motivated embeddings allow to face issues of theoretical consistency, like unitarity, which might be already relevant at the energy ranges currently experimentally probed~\cite{Kahlhoefer:2015bea}.

\noindent
This work is focused on the case of a dirac fermion DM interacting with SM fermions through a spin 1 mediator. The mediator is interpreted as the gauge boson of a new $U(1)^{'}$ symmetry emerging, together with the SM gauge group, from the breaking of GUT symmetries like $SO(10)$ and $E6$.

\section{The models}   
In this work we will consider 3 GUT inspired $Z^{'}$ realizations. Two of them, labeled as $E6_\chi$ and $E6_\eta$ are inspired by embedding in $E6$. The last realization, labeled as LR, is instead based on a $SO(10)$ embedding (see~\cite{Langacker:2008yv} for a review on the theoretical origin). For comparison we will also include in our analysis a 'bottom-up' inspired scenario, i.e. the SSM in which the $Z^{'}$ has the same couplings with SM fermions as the SM $Z$ boson. The couplings of the $Z^{'}$ with SM fermions are fixed by the breaking pattern of the GUT group. The corresponding values for the considered $Z^{'}$ realizations can be found for example in~\cite{Han:2013mra,Arcadi:2015nea}. In order to achieve a framework fully compatible with UV complete models the couplings of the DM with the $Z^{'}$ should be, similarly, fixed. This point is however still under investigation and will be left to future work (see e.g.~\cite{Nagata:2015dma} for some already existing concrete examples.). We will thus still regard the DM couplings as free parameters and examine possible indications, to their structure, provided by experimental constraints.

\noindent
The relevant interactions are encoded in the following lagrangian:
\begin{equation}
\label{eq:lagrangian}
\mathcal{L}=g_d V_\chi \bar{\chi} \gamma^{\mu}\left(1-\alpha \gamma_5\right) \chi Z^{'}_{\mu}+ g_d \sum_f \bar f \gamma^\mu \left(V_f - A_f \gamma_5\right) f Z^{'}_{\mu}
\end{equation}
where, for phenomenological reasons, we have adopted vector and axial currents rather than left-handed and right-handed ones. As already mentioned we will consider a total of 4 definite set of assignations for the couplings $V_f, A_f$\footnote{For simplicity we are assuming only a direct coupling of the $Z^{'}$ with SM fermions. In a complete framework additional couplings from mass and kinetic mixing of the $Z^{'}$ with the SM $Z$-boson are in general present, leading to additional interactions of the DM with SM gauge bosons. For simplicity we are neglecting this last feature leaving this possibility to future studies.}, while the coupling $g_d$ coincides with the SM gauge coupling $g$ in the case of the SSM and with $g_{\rm GUT}=\sqrt{5/3}\tan\theta_{\rm W} g$ in the other cases. We have introduced, for the case of DM interactions, the parameter $\alpha=\frac{A_\chi}{V_\chi}$. In the scenario presented in this work the coupling $V_\chi$ results strongly constrained by the limits from Dark Matter Direct Detection (DD) and does not give a sizable contribution to the relevant observables. Our setup has then three fundamental parameters, being the DM mass $m_\chi$, the $Z^{'}$ mass $m_{Z^{'}}$ and $\alpha$ itself.

\section{Direct Detection}

The $Z^{'}$ realizations illustrated in the previous section feature both unsuppressed Spin Independent (SI) and Spin Dependent (SD) direct detection cross-sections. The corresponding expressions (for the scattering on protons) are, schematically, given by:
\begin{equation}
\label{eq:cross_DD}
\sigma_{\chi p}^{\rm SI}=\frac{g_d^4 \mu_{\chi}^2}{\pi m_{Z^{'}}^4} |V_\chi|^2 \alpha_{\rm SI},\,\,\,\,\sigma_{\chi p}^{\rm SD}=\frac{3 g_d^4 \mu_\chi^2}{\pi m_{Z^{'}}^4} \alpha^2 |V_\chi|^2 \alpha_{\rm SD}
\end{equation}
where the coefficients $\alpha_{\rm SI}$ and $\alpha_{\rm SD}$ depend on the couplings of the $Z^{'}$ with SM fermions and the target nuclei of the DM experiment considered for eventual comparisons~\cite{Arcadi:2013qia,Arcadi:2014lta}.

\begin{figure}[htb]
\begin{center}
\subfloat{\includegraphics[width=6.5 cm]{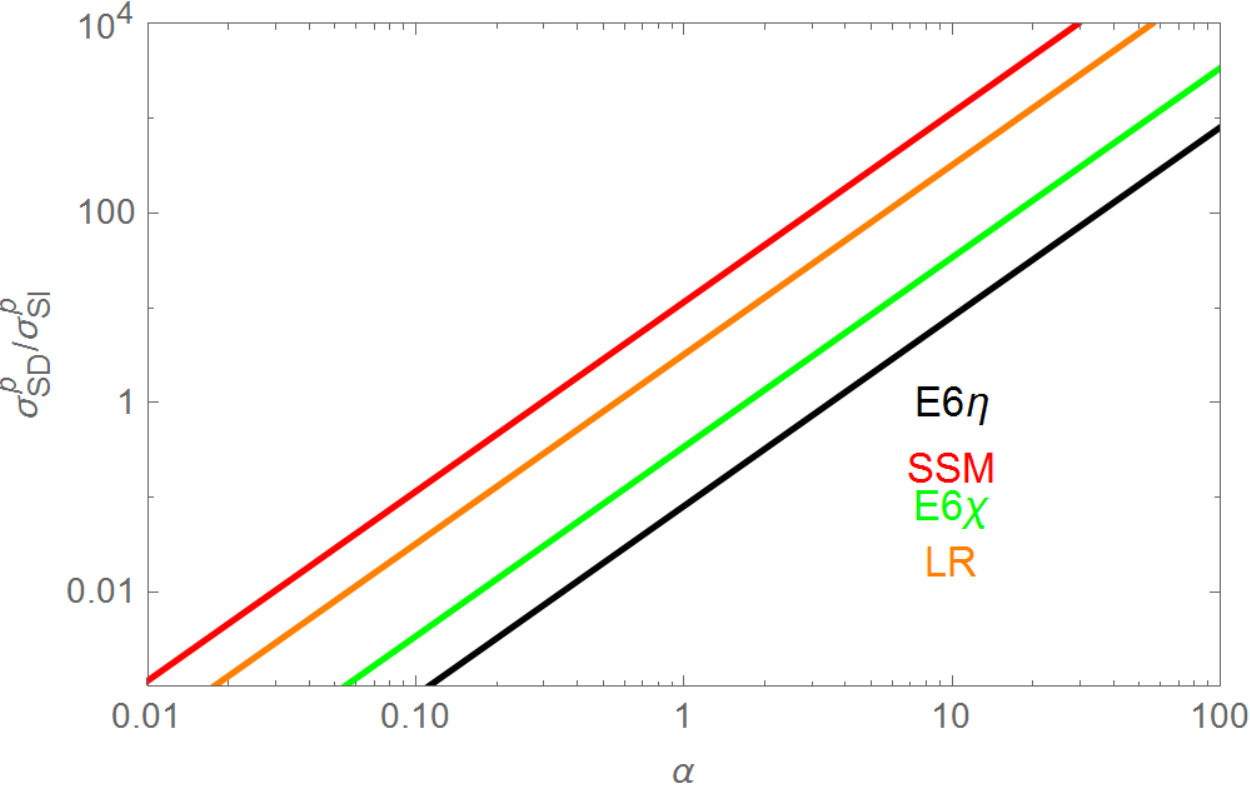}}
\subfloat{\includegraphics[width=6.2 cm]{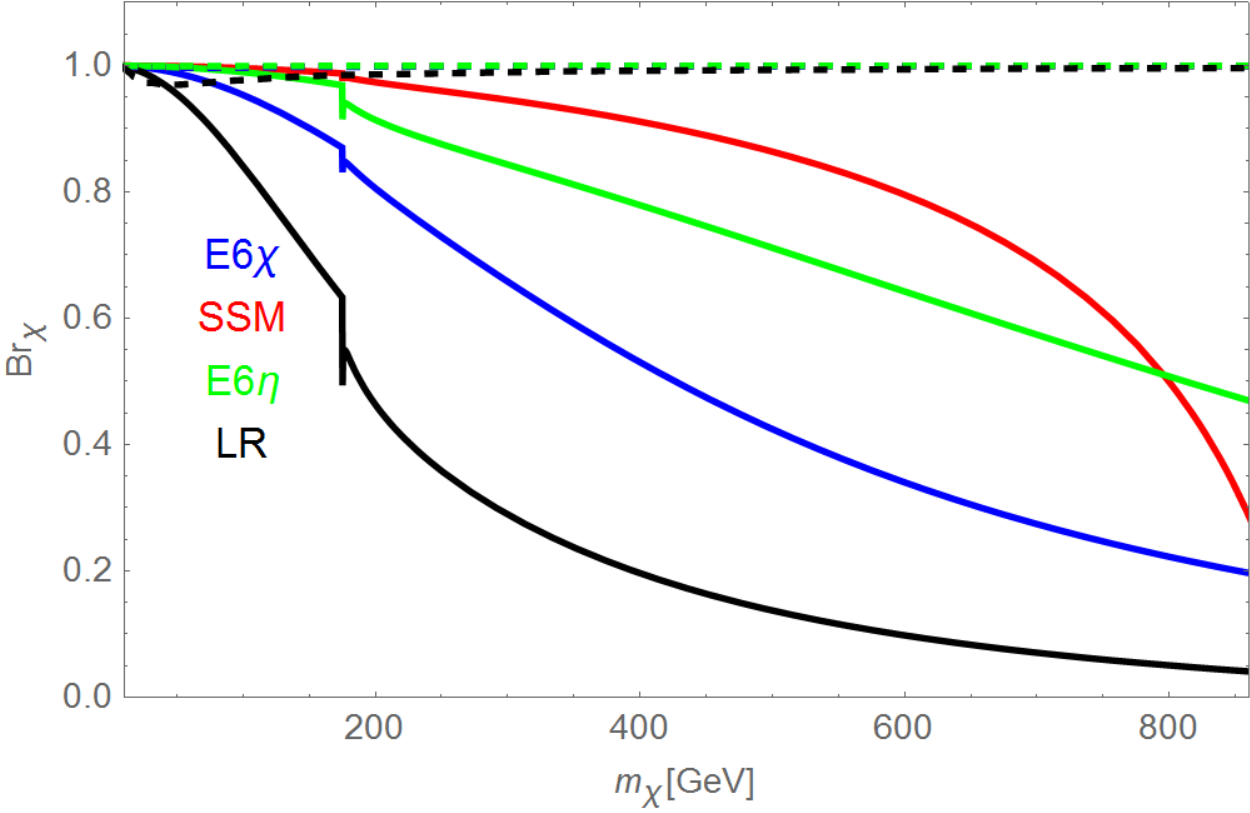}}
\caption{\footnotesize{Left Panel: Ratio of the SD cross section of the DM on proton over the SI component, as function of the parameter $\alpha$ for the $Z^{'}$ realizations considered. Right Panel:Invisible branching fraction, as given by eq.~(\ref{eq:brinv}), of the $Z^{'}$, as function of the DM mass, for the realizations considered in this work. Solid (dashed) lines refer to the case in which the Invisible branching fraction is set by the DM relic density (limit by PICO).}}
\label{fig:DD}
\end{center}
\end{figure}

\noindent
Interestingly, the ratio of the two components of the cross-section is independent from the mass of the $Z^{'}$ and it is set by the value of $\alpha$ and the ratio $\alpha_{\rm SI}/\alpha_{\rm SD}$. The ratio $\sigma_{\chi p}^{\rm SD}/\sigma_{\chi p}^{\rm SI}$, as function of $\alpha$, is reported, for the considered $Z^{'}$ realizations, in the left panel of fig.~(\ref{fig:DD}). From this it is straightforward to argue that, complete frameworks, such that the value of $\alpha$ is as well a theoretical prediction, could be potentially distinguished by a combined detection of the two components, i.e. SI and SD, of the DM scattering cross-section of nuclei.

\noindent
Another interesting correlation can be established with the requirement of the correct DM relic density. Indeed, in our parametrization, the DM pair annihilation cross-section into SM fermions can be expressed as:
\begin{equation}
\label{eq:cross_relic}
\langle \sigma v \rangle= \frac{m_\chi^2}{\pi m_{Z^{'}}^4} |V_\chi|^2 \left[\left(a_{V}+b_{V}v^2\right)+\alpha^2 \left(a_{A}+b_{A}v^2\right)\right]
\end{equation}
where we have employed the conventional velocity expansion which is valid away from resonances (as clarified in the next subsection this assumption is safe at least for DM masses below few hundreds of GeV). Being the value of the coupling $|V_\chi|$ severely constrained by the limits on the SI scattering cross-section from the LUX experiment, the correct relic density is achieved, for each assignation of the pair $(m_\chi,m_{Z^{'}})$, for a definite value of $\alpha$. We can then formulate the prediction for the SD DM scattering cross-section in terms of the DM pair annihilation cross-section. This prediction, at least within the range of validity of the velocity expansion, depends on $\alpha$ and the DM mass but it is independent from $m_{Z^{'}}$. Moreover it can, in turn, be confronted with current experimental limits or, possibly, the sensitivity of future experiments. For the $Z^{'}$ realizations considered here, current constraints~\cite{Amole:2015lsj} can already probe DM masses below 100-150 GeV~\cite{Arcadi:2015nea}.

\section{Collider prospects} 

The results discussed in the previous section have as well profound implications in the prospects for collider searches. In general the kind of setup discussed offers a rather broad phenomenology, being in general accessible signals associated to both the ``invisible'' (i.e. in a DM pair) and the ``visible'' (i.e. in SM particles) decay channels of the $Z^{'}$. The very interesting possible complementarity between this two kind of signals is discussed, for example, extensively in~\cite{Alves:2015pea,Chala:2015ama,Alves:2015mua}. We will instead focus here on a rather definite aspect, along the same lines as~\cite{Arcadi:2013qia}, being searches of dilepton resonances. This kind of searches are one of the customary tools for probing the existence of new gauge bosons and already severe limits has been already imposed, excluding masses of the $Z^{'}$ of the order of 2.5-3 TeV~\cite{Aad:2014cka}. However, this kind of results rely on the assumption that the $Z^{'}$ has only decay channels into SM states. In the case, instead, that it is also coupled with a pair of DM states, the production cross-section of the dilepton resonance is rescaled according the invisible branching fraction $Br_\chi$ of the $Z^{'}$ as: 
\begin{equation}
\label{eq:brinv}
\sigma_{Z'll}\to \left({g_d\over g}\right)^2 \times (1-Br_\chi) \times \sigma_{Z'll},\,\,\,\,\,Br_\chi={\left[ 1+  \left( \frac{2 g_d^2 \mu_{\chi N}}{m_{Z'}^2 \sqrt{\pi}} \right)^2
\frac{\sum_f n_f[ |V^{'}_f|^2  + |A^{'}_f|^2]}
{\sigma_{\chi N}^{\rm{SI}}/\alpha_{\rm{SI}} +1/3 \sigma_{\chi N}^{\rm{SD}}/ \alpha_{\rm{SD}} }   \right]}^{-1} 
\end{equation}
where the sum runs over the SM fermions, with $n_f$ being the color factor. Given the very different experimental limits, the dominant contribution is from the SD cross-section. Interestingly, we can use the results discussed in the previous subsection to relate the limits from DD and, possibly, the correct relic density to the LHC limits from dilepton resonances.
\noindent
We show on the right panel of fig.~(\ref{fig:DD}) the values of the invisible branching fractions in the different $Z^{'}$ models allowed by the correct DM relic density (solid lines) and by current limits on the DD cross-section (dashed lines). As evident, the latters still allow for a substantially invisible $Z^{'}$. Dark matter relic density, on the contrary, provides sizable constraints, expecially in high DM region, with sizable difference between the various realizations, possibly implying different detection prospects.

\begin{figure}[htb]
\begin{center}
\includegraphics[width=6.5 cm]{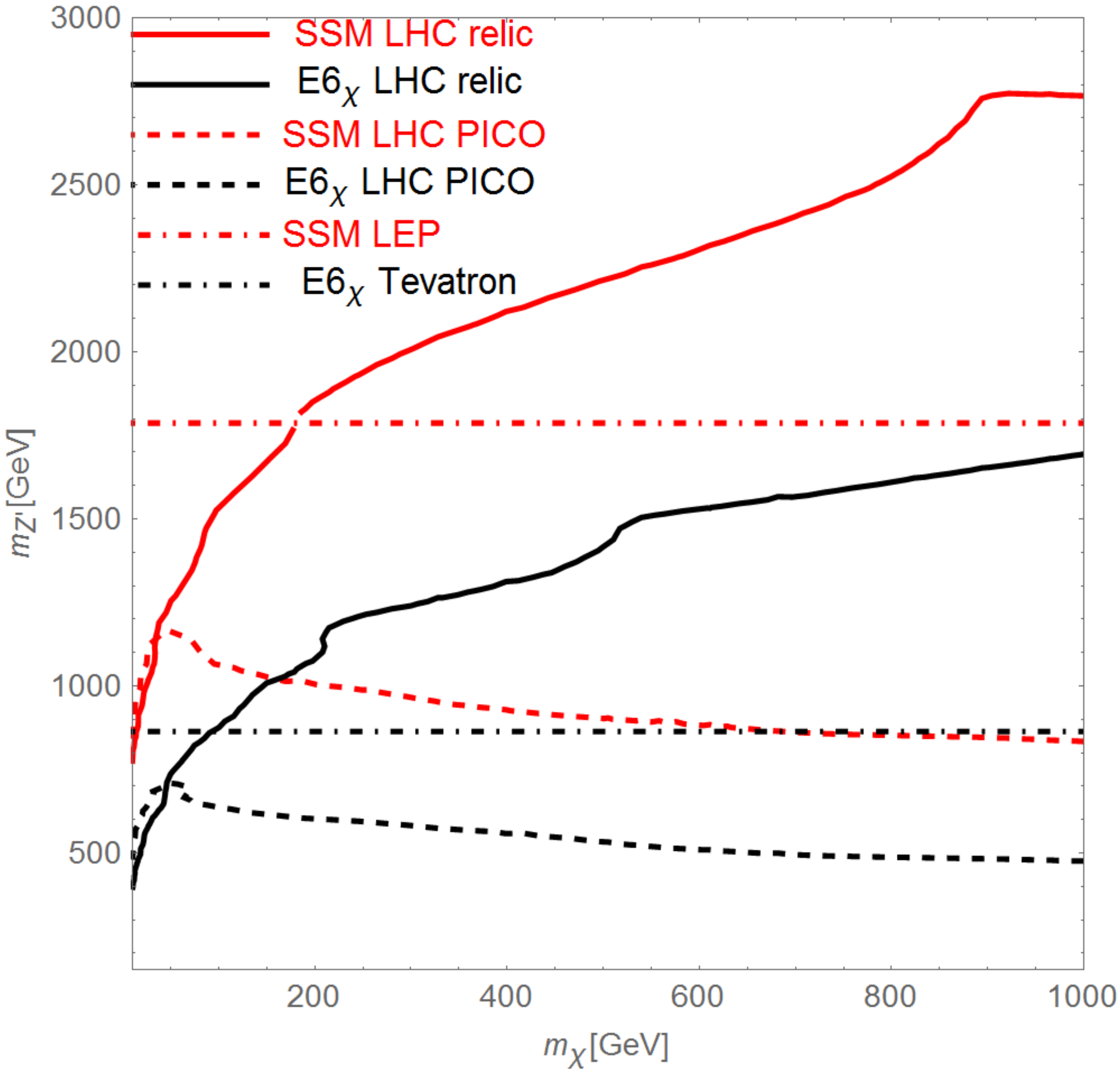}
\caption{\footnotesize{Reformulation of the limits from searches of dilepton resonances, in the bidimensional plane $(m_\chi,m_{Z^{'}}$, in presence of non-null invisible branching fraction of the DM for two $Z^{'}$ realizations, namely the SSM (red lines) and the $E6_\chi$ model (black lines). Solid lines refer to the value of the invisible branching fraction compatible with the thermal relic density while the dashed lines correspond to the maximal value of the invisible branching fraction allowed by PICO experiment. The dot-dashed curves represent the limits by LEP/Tevatron.}}
\label{fig:LHC}
\end{center}
\end{figure}

\noindent
As an example, fig.~(\ref{fig:LHC}) shows the limits from dilepton searches in the bidimensional plane $(m_\chi,m_{Z^{'}})$, for two $Z^{'}$ realizations, once a non null $Br_\chi$ is taken into account. Apart the region at $m_\chi \lesssim 100-150\,\mbox{GeV}$ the limits obtained when $Br_\chi$ is governed by the DM relic density (solid lines) are more stringent than the one given by experimental limits on DD (dashed lines). In particular we have than in the case of the SSM the allowed invisible branching fraction at high DM masses is negligible and then, consistently, the limit converge to the one originally quoted by experimental collaboration, which does not depend on the DM mass. Even in presence of a high invisible branching fraction the limit on the $Z^{'}$ mass cannot arbitrarily decrease because of limits from LEP and Tevatron on $m_{Z^{'}}$~\cite{Langacker:2008yv}, sensitive to the off-shell production of the $Z^{'}$ and not influenced by the reduction due to the invisible branching fraction, occurring only at the s-channel resonance.

\section{Conclusions and outlook}

Simplified models are a very powerful tool for combining different search strategies of new physics. The complementarity among different searches is further enforced by limiting to the subset of simplified models which can be embedded in theoretically motivated particle physics frameworks. Indeed in such a case it is possible to reduce the number of free parameters, having predictions of the relevant couplings, and to go beyond the comparison with current experimental constraints possibly inferring new signals which can be tested at next future facilities. This work has discussed some examples, still sticking on a rather simplified setup, of a very broad phenomenology which will be explored in further studies.

\noindent
{\bf Acknowledgments}:The author warmly thanks the organizers of the TAUP conference for the chance of giving this contribution.
\noindent
The author is supported by the ERC advanced Grant Higgs@LHC.

\section*{References}

\bibliography{iopart-num}{}
\bibliographystyle{iopart-num}
\end{document}